\documentclass[11pt]{article}
\usepackage{amsmath}
\usepackage{amssymb}
\newtheorem{example}{Example}[section]

\newtheorem{note}{Note}[section]

\begin{document}
\begin{center}
{\LARGE\bf The Distribution and Quantiles of  Functionals of}\\[1ex]
{\LARGE\bf Weighted Empirical Distributions when}\\[1ex]
{\LARGE\bf Observations have Different Distributions}\\[1ex]
by\\[1ex]
Christopher S. Withers\\
Applied Mathematics Group\\
Industrial Research Limited\\
Lower Hutt, NEW ZEALAND\\[2ex]
Saralees Nadarajah\\
School of Mathematics\\
University of Manchester\\
Manchester M13 9PL, UK
\end{center}
\vspace{1.5cm}
{\bf Abstract:}~~This paper extends Edgeworth-Cornish-Fisher expansions for
the distribution and quantiles of nonparametric
estimates in two ways. Firstly it allows  observations to have
 different distributions. Secondly it allows  the  observations to be
weighted in a predetermined way.
The use of weighted estimates has a long history including applications to
regression, rank statistics and Bayes theory. However asymptotic
results have generally
been only first order (the CLT and weak convergence). We give third order
asymptotics for the distribution and percentiles of any smooth functional
 of a weighted empirical distribution,
thus allowing a considerable increase in accuracy over earlier CLT results.

 Consider independent non-identically distributed ({\it non-iid})
observations $X_{1n}, \ldots, X_{nn}$
in $R^s$. Let $\widehat{F}(x)$ be their {\it weighted empirical
distribution} with weights $w_{1n}, \ldots, w_{nn}$.
 We obtain cumulant expansions and hence Edgeworth-Cornish-Fisher expansions for
$T(\widehat{F})$ for any smooth functional $T(\cdot)$ by extending
the concepts of von Mises derivatives to signed measures of total measure 1.
As an example we give the cumulant  coefficients needed for  Edgeworth-Cornish-Fisher expansions to  $O(n^{-3/2})$ for the sample variance
when observations are non-iid.


\section{Introduction and Summary}

Withers (1983, 1988) gave third order asymptotics for the distribution of
functionals of empirical (or sample) distributions for iid observations.
This paper extends these results to non-iid weighted observations.

Traditional inference is based on the empirical distribution function.
This gives each observation equal weight. However
 in many contexts it is more appropriate to weight the observations
differently. An important class of weighted statistics are the rank
statistics studied by Hajek and Sidak (1967).
They gave first order (asymptotic) results
both for iid observations, and for the {\it contiguous} case
where the observations are from distributions approaching the null
case. However they did not deal with the case where observations have fixed distinct distributions.
A seminal contribution to the theory of weighted empirical distributions
was made by
Koul (1992) who gave first order properties for linear models.
This was extended by another seminal contribution, Koul (2002), to allow for random weights with
applications to $M$- and $R$- estimates as well as to autoregressive processes.
However, he confined his focus to first order (weak convergence) results.
Optimality of  certain weights in a Bayesian setting
was proved by Chernoff and Zacks (1964) for testing the hypothesis of a jump in the mean.
For other work, including a comprehensive account of the literature,
we refer the readers to Lahiri (1992a, 1992b, 1992c) and Lahiri (2003).

This paper - following on from Withers and Nadarajah (2008) -
gives cumulant expansions - and hence Edgeworth-Cornish-Fisher
expansions - for smooth functionals of weighted empirical distributions
for {\it arbitrary} non-iid observations.
These cumulant expansions are given in Section 2.
As simple examples, Section 3 applies these results to the mean and variance.
Section 4 gives a chain rule for the functional derivative of a function
of several functionals, and uses this to obtain the leading cumulant
coefficients for the Studentized mean,
and the coefficient of variation.
For completeness the Edgeworth-Cornish-Fisher expansions for the distribution
and quantiles are given in
Appendix A, as well as the multivariate Edgeworth expansion.

Let $X_{1n}, \ldots, X_{nn}$ be independent random variables in $R^s$ with distributions $F_{1n}, \ldots, F_{nn}$.
Let  $w_{1n}, \ldots, w_{nn}$ be given real numbers adding to $n$:
\begin{eqnarray}
w_{1n} + \ldots + w_{nn} = n.\label{w}
\end{eqnarray}
The simplest example giving more weight to the later observations are the weights $w_{in}=2i/(n+1)$
shown by Chernoff and Zacks (1964) to be optimal in a Bayesian setting
 for testing for a jump in the mean.
The mean with these weights was also used by Kander and Zacks (1966) and others.

Define the {\it weighted empirical distribution} as
\begin{eqnarray*}
\widehat{F}(x) = n^{-1} \sum^n_{i=1} w_{in} I(X_{in} \leq x)
\end{eqnarray*}
for $x$ in $R^s$, where $I(A) = 1$ or 0 for $A$ true or false.
Its mean is
\begin{eqnarray}
F(x) = E\ \widehat{F}(x) = n^{-1} \sum^n_{i=1} w_{in} F_{in}(x).
\label{hash.eg3}
\end{eqnarray}
$F$ has moments
\begin{eqnarray*}
m_r=m_r(F)=\int x^rdF(x)=n^{-1} \sum^n_{i=1} w_{in} m_r(F_{in}),
\
\mu_r=\mu_r(F)= E_F (X-m_1)^r
\end{eqnarray*}
with sample versions
\begin{eqnarray*}
\widehat{m}_r=m_r(\widehat{F})=n^{-1} \sum^n_{i=1} w_{in} X_{in}^r,
\
\widehat{\mu}_r=\mu_r(\widehat{F})=n^{-1} \sum^n_{i=1} w_{in} (X_{in}-\widehat{m}_1)^r.
\end{eqnarray*}
For convenience {\it we now suppress the subscript $n$}
 and write $X_{i}=X_{in}$, $F_{i}=F_{in}$, $w_{i}=w_{in}$.

Let $T(G)$ be a smooth real functional defined for all signed
measures $G(x)$ on $R^s$ with total measure $G(\mathbf{\infty}$) = 1.
(This is the condition that requires the constraint (\ref{w}).)
The $r$th order (von Mises) functional derivative of $T(G)$
$T_G (x_1,  \ldots, x_r)$ for $x_1, \ldots, x_r$ in $R^s$
may be defined just as in the case when $F(x)$ is a probability
distribution, and the von Mises-Taylor expansion for two such
signed measures $G$, $H$ remains valid, giving an iterative method for
obtaining higher derivatives:
\begin{eqnarray}
T(G) - T(H) = \sum^\infty_{r=1} T_r (G,H) / r!,
\label{1.4}
\end{eqnarray}
where
\begin{eqnarray*}
T_r (G,H) = \int \ldots \int T_H (x_1, \ldots, x_r) dG(x_1) \ldots dG(x_r),
\end{eqnarray*}
where $T_H (x_1, \ldots, x_r)$ is made unique by two conditions:
first that it is symmetric under permutation of arguments $x_1, \ldots, x_r$;
and second that it satisfies
\begin{eqnarray}
\int T_F (x_1, \ldots, x_r) dF(x_1)\equiv 0
\label{uniq}
\end{eqnarray}
for $r\geq 1$.
The first derivative,
$T_H(x)$ is just the coefficient of $\epsilon$ in the Taylor expansion of
$T(H+\epsilon(\delta_x-H))$ about $\epsilon=0$, where $\delta_x$ is the distribution putting weight 1 at $x$. It
 is sometimes called the {\it influence function} of $T(F)$.
 The rule for differentiating $S(H) = T_H (x_1, \ldots, x_r)$ given
by Theorem 2.1 of Withers (1983) remains valid:
\begin{eqnarray}
T_H(x_1\ldots x_{r+1}) = S_{H}(x_{r+1})
+ \sum^r_{i=1}T_H\left< {x_1 \ldots x_{r+1}}\right>_i,
\label{diff}
\end{eqnarray}
where $S(H) = T_H ({x_1 \ldots x_{r}})$
and $\langle \cdot \rangle_i$ means ``drop the $i$th column''.
For example, putting $r=1$, the second derivative is given by
$T_H(x_1x_2) = S_{H}(x_2) +T_H(x_2)$, where $S(H)=T_H(x_1)$.
The theory of statistical functionals was pioneered by von Mises (1947).
 The importance and use of the influence function has been widely
used to obtain the asymptotic variance of general estimates,
\begin{eqnarray*}
n var(T(\widehat{F})) \rightarrow \int T_F(x)^2dF(x)
\end{eqnarray*}
as $n\rightarrow \infty$.
The  second derivative has been used to estimate and correct for bias:
\begin{eqnarray*}
E\ T(\widehat{F}) -T(F) =n^{-1}\int T_F(x,x)dF(x)/2 +  O(n^{-2}).
\end{eqnarray*}
This was used by Jaeckel (1972) to justify the infinitesimal jackknife.
However the use of other
higher order derivatives has not been widespread.
The reason for
seems to be that not until Withers (1983), was the formula (\ref{diff})
available to compute higher order derivatives. Nevertheless their use since then
has been disappointing. Perhaps this is due to a common misconception that
for $T(F)$ a function of moments, it is just as easy to simply use ordinary
Taylor expansions. To see that this is not true, consider the following
simple example.

\begin{example}
Let us compute the asymptotic
variance of the $r$th central sample moment, that is, $a_{21}/n$, by both
the functional method and the ordinary Taylor expansion method, when
$T(F)=\mu_r=\int (x-\mu)^rdF(x),\ \mu=m_1(F)$, observations are iid, and
the ordinary unweighted empirical distribution is used.
Then
\begin{eqnarray}
&&
T_x = (x-\mu)^r-\mu_r-r(x-\mu)\mu_{r-1},
\nonumber
\\
&&
a_{21} = \int T_x^2 dF(x)
=\mu_{2r}-\mu_r^2-2\mu_{r-1}\mu_{r+1}+r^2\mu_{r-1}^2\mu_2.
\label{ex}
\end{eqnarray}
The ordinary Taylor series method writes $\mu_r$ as a function of the non-central moments:
\begin{eqnarray*}
\mu_r=\sum_{i=0}^r (-1)^{r-i} {r\choose i} m_i m_1^{r-i}
=(-1)^{r-1}(r-1)m_1^r +\sum_{i=2}^r (-1)^{r-i}{r\choose i}m_im_1^{r-i}
=t(m_1,\cdots,m_r)
\end{eqnarray*}
say, with derivatives $t_{\cdot i}=\partial t(m_1,\cdots,m_r)/\partial m_i$
given by
\begin{eqnarray*}
t_{\cdot 1}=(-1)^{r-1}(r-1)rm_1^{r-1} +\sum_{i=2}^r (-1)^{r-i}{r\choose i}
(r-i)m_im_1^{r-i-1},
\end{eqnarray*}
and for $1<i\leq r$,
\begin{eqnarray*}
t_{\cdot i}=(-1)^{r-i}{r\choose i}m_1^{r-i}.
\end{eqnarray*}
Also $covar(m_i(\widehat{F}),m_j(\widehat{F}))\approx (m_{i+j}-m_im_j)/n$.
So,
\begin{eqnarray*}
a_{21}=\sum_{i,j=1}^r t_it_j\ (m_{i+j}-m_im_j).
\end{eqnarray*}
The challenge to these advocates of the ordinary Taylor method is to show
that this reduces to (\ref{ex}).
Even for the variance, this
takes some time.
\end{example}

In Section 2 we use the von-Mises expansion for $\widehat{\theta}=T(\widehat{F})$ to obtain the basic cumulant expansion
\begin{eqnarray}
\kappa_r (\widehat{\theta}) = \sum^\infty_{j = r-1} a_{rj} n^{-j}
\label{hash.eg5}
\end{eqnarray}
for $r \geq 1$ needed for the Edgeworth-Cornish expansions of Appendix A.
(So, $a_{21}/n$ and $a_{11}/n$ are the asymptotic variance and bias of
$T(\widehat{F})$, viewed as an estimator of $T({F})$.)
These expansions require that the
cumulant coefficients $\{a_{rj}\}$ are all bounded as $n \to \infty$.
This is true if
\begin{eqnarray}
W_r  = n^{-1} \sum^n_{i=1} w^r_i
\label{Wr}
\end{eqnarray}
is bounded for $r\geq 1$ and the $[\cdot ]_{ij\ldots}$ functions of Section 2 are bounded.
The expansions of Appendix A also require that $a_{21}$ is bounded away from 0.
Typically this is true if $W_2$  is bounded away from 0.
The first order results of Koul (2002)
 may be reconciled with
our first order results by noting that he works with $d_i=w_i/(nW_2)^{1/2}$.
Throughout, we assume that all weights $\{w_{in}\}$ are bounded and that all weight functions have finite derivatives.

\section{Moment and Cumulant Expansions for Estimates}
\addtocounter{section}{0}
\setcounter{equation}{0}

What happens to non-parametric estimates when the assumption that observations
come from the same distribution breaks down?
Here we derive the cumulant expansion (\ref{hash.eg5}) for $\widehat{\theta}=T(\widehat{F})$,
giving explicitly the cumulant coefficients of (\ref{hash.eg5}),
$a_{21}$, $a_{11}$, $a_{32}$, $a_{22}$, $a_{43}$,
needed for third order expansions and inference for non-iid observations.

We follow the approach of Withers (1983) deriving the cumulant expansion from the moment expansion
\begin{eqnarray*}
E\  \{ T(\widehat{F}) - T(F) \}^r = \sum_{j\geq r/2} a^\prime_{rj} n^{-j}
\end{eqnarray*}
for $r \geq 1$ using the relations between their coefficients
$\{a_{rj} \}$ and $\{ a^\prime_{rj} \}$ given in Theorem 3.1 of Withers (1983).
We use the following notation with $F$ of (\ref{hash.eg3}):
\begin{eqnarray*}
T_{x_1 \ldots x_r}
&=&
T_F(x_1, \ldots, x_r),
\\
\left[1^r \right]_i
&=&
E\ T^r_{X_i}= \int T^r_{x_1} dF_i (x_1),
\\
\left[ 1^r,{11}\right] _i
&=&
E\ T^r_{X_i}T_{X_i X_i}  = \int T^r_{x_1} T_{x_1 x_1} dF_i (x_1),
\\
\left[ 1^r, 12^s, 2^t\right]_{ij}
&=&
E^{ind}\ T^r_{X_i}T_{X_i X_j}^s T^t_{X_j}
= \int \int T^r _{x_1} T^s_{x_1 x_2} T^t _{x_2} dF_i(x_1) dF_j (x_2),
\\
\left[ 1,122\right]_{ij}
&=&
E^{ind}\ T_{X_i}T_{X_i X_jX_j} = \int\int T_{x_1} T_{x_1x_2x_2} dF_i (x_1) dF_j(x_2),
\\
\left[ 1,2,3,123\right]_{ijk}
&=&
E^{ind}\ T_{X_i} T_{X_j}T_{X_k}T_{X_i X_jX_k}
\\
&=&
\int\int\int T_{x_1} T_{x_2} T_{x_3} T_{x_1x_2x_3} dF_i (x_1)dF_j(x_2)dF_k(x_3)
\end{eqnarray*}
and so on, where $E^{ind}$ means $E$ treating $X_i,\ X_j,\ \cdots$ as independent. Now set
\begin{eqnarray*}
&&
\left[ 1^r\right] = n^{-1} \sum^n_{i=1} w^r_i \left[ 1^r\right]_i,
\\
&&
\left[1^r,{11}\right] = n^{-1} \sum^n_{i=1} w^{r+2}_i \left[ 11\right]_i,
\\
&&
\left[1^r,{12}^s, 2^t\right] = n^{-2} \sum^n_{i,j=1} w^{r+s}_i  w_j^{s+t} \left[1^r, 12^s, 2^t
\right]_{ij},
\end{eqnarray*}
and more generally for $S$ a number of sequences from and including $1,2,\ldots, r$ we set
\begin{eqnarray*}
[S] = n^{-r} \sum^n_{i_1=1} \ldots \sum^n_{i_r =1}
w^{\lambda_1}_{i_1} \ldots w^{\lambda_r}_{i_r} [S]_{i_1 \ldots i_r},
\end{eqnarray*}
where $\lambda_j$ is the number of times $j$ occurs in $S$.

In Appendix B we derive the following expressions for the cumulant coefficients
needed for the Edgeworth-Cornish-Fisher expansions of $T(\widehat{F})$ to $O(n^{-3/2})$:
\begin{eqnarray}
&&
a_{10} = T(F),
\quad
a_{21} = \{ 1^2\},
\label{hash.eg16}
\\
&&
a_{11} = \{ 11\}/2 \mbox{ and } a_{32} = \{ 1^3\} + 3 \{ 1, 2, 12\},
\label{hash.eg17}
\\
&&
a_{22} = \{ 1, 11 \}
   + \{ 12^2\}/2
   + \{ 1, 122 \},
\label{hash.eg18}
\\
&&
a_{43} = \{ 1^4\}
  - 3\{ 1^2, 1^2\}
  + 12 \{ 1, 2^2, 12 \}
  + 12 \{ 1, 2, 13, 23 \}
  + 4 \{ 1, 2, 3, 123 \},
\label{hash.eg19}
\end{eqnarray}
where
\begin{eqnarray*}
\{1^r\}
&=&  n^{-1} \sum^n_{i=1} w^2_i \mu_r(T_F(X_i))
\\
&=&
\begin{cases}
 n^{-1} \sum^n_{i=1} w^2_i ( [1^2]_i - [1]^2_i ), & \text{if } r=2,\\
 n^{-1} \sum^n_{i=1} w^3_i \{ [ 1^3 ]_i  - 3 [ 1 ]_i [ 1^2 ]_i   + 2 [ 1 ] ^3_i \}, & \text{if }r=3,\\
 n^{-1} \sum^n_{i=1} w^4_i \{ [ 1^4]_i  - 4 [ 1 ]_i [ 1^3 ]_i [ 1^3 ] _i   + 6 [ 1 ] ^2_i [ 1^2 ]_i  - 3[ 1] ^4_i \},  & \text{if }r=4,
\end{cases}
\\
\{ 11\}
&=& \left[ 11\right]
   - n^{-1}\sum^n_{i=1} w^2_i [ 12 ]_{ii}
   = n^{-1}\sum^n_{i=1} w^2_i \{ [ 11 ]_i
     - [ 12]_{ii} \},
\nonumber
\\
\quad
\{ 1, 2, 12 \}
&=&
n^{-2} \sum^n_{i,j=1} w^2_i w^2_j ( [ 1, 2, 12 ]_{ij}
   - 2 [ 1 ]_j [ 1, 12 ]_{ij}
   +  [ 1]_i [ 1 ]_j [ 12 ]_{ij} ),
\nonumber
\\
\{ 1, 11 \}
&=&
n^{-1} \sum^n_{i=1} w^3_i ( [ 1, 11 ] _i
  - [ 1 ]_i [ 11 ]_i
  - 2 [ 1, 12 ]_{ii}
  + 2 [ 1]_i [ 12 ] _{ii} ),
\nonumber
\\
\{ 12^2 \}
&=&
n^{-2} \sum^n_{i,j=1} w^2_i w^2_j ( [ 12^2 ]_{ij}
  - 2 [ 12, 13 ]_{ijj}
  + [ 12 ]^2_{ij} ),
\nonumber
\\
\{ 1, 122 \}
&=&
n^{-2} \sum^n_{i,j =1} w^2_i w^2_j ( [ 1, 122 ]_{ij}
  - [ 1, 123 ]_{ijj}
  - [ 1 ]_i [ 122 ]_{ij}
  + [ 1 ]_i [ 123 ] _{ijj} ),
\nonumber
\\
\{ 1^2, 1^2 \}
&=&
n^{-1} \sum^n_{i=1} w^4_i \mu_2(T_F(X_i))^2
\ =\ n^{-1} \sum^n_{i=1} w^4_i ( [ 1^2 ]_i  - [ 1] ^2_i )^2,
\nonumber
\\
\{ 1, 2^2, 12 \}
&=&
n^{-2} \sum^n_{i,j=1} w^2_i  w^3_j
   \left(  [ 1, 2^2, 12]_{ij}
  - [ 1^2 ]_j
    [ 1, 12 ]_{ij}
  -  [ 1 ]_j [ 1, 2, 12]_{ij} \right.
\nonumber
\\
& &
\left. + 2 [ 1 ]^2_j [ 1, 12]_{ij}
  - [ 1 ] _i \tau_{ij} \right),
\nonumber
\\
\tau_{ij}
&=&
[ 1^2, 12 ]_{ji}
  - [ 1^2 ]_j   [ 12 ]_{ij}
  - [ 1]_j [ 1, 12 ]_{ji}
  + 2[ 1 ] ^2_j [ 12] _{ij},
\nonumber
\\
\{ 1, 2, 13, 23 \}
&=&
n^{-3} \sum^n_{i,j,k=1} w^2_i w^2_j w^2_k
  (
  [ 1, 2, 13, 23 ] _{ijk}
  - [ 1 ]_k [ 1, 12, 23 ]_{ijk}
\nonumber
\\
& &
- [ 1, 12 ]_{ij}[ 1, 12 ]_{kj}
  + 2[ 1 ]_k [ 12 ]_{jk}[ 1, 12 ]_{ij}
  - [ 1 ]_i [ 1 ]_k [ 12]_{ij}
    [ 12]_{jk} ),
\nonumber
\\
\{ 1, 2, 3, 123 \}
&=&
n^{-3} \sum^n_{i,j,k=1} w^2_i w^2_j w^2_k
    ( [ 1, 2, 3, 123] _{ijk}
  - 3[ 1 ] _k
    [ 1, 2, 123 ] _{ijk}
\nonumber
\\
&&
  + 3[ 1 ]_j [ 1 ]_k [ 1, 123 ] _{ijk}
  - [ 1 ]_i [ 1]_j [ 1 ]_k
    [ 123 ]_{ijk} ).
\nonumber
\end{eqnarray*}
In addition the following coefficient is useful for the calculation of
the second order bias:
\begin{eqnarray}
a_{12} = \{ 111\} /6 + \{ 1122 \}/8,
\label{hash.eg20}
\end{eqnarray}
where
\begin{eqnarray*}
\{ 111 \} = n^{-1} \sum^n_{i=1} w^3_i
  \left( \left[ 111 \right]_i
  - 3\left[ 122 \right]_{ii}
  + 2\left[ 123 \right]_{iii} \right)
\end{eqnarray*}
and
\begin{eqnarray*}
\{ 1122 \} = n^{-2} \sum^n_{i,j=1} w^2_i w^2_j
  \left( \left[ 1122 \right]_{ij}
  - 2 \left[ 1233 \right]_{iij}
  + \left[ 1234 \right] _{iijj} \right).
\nonumber
\end{eqnarray*}
Typically $a_{21}$ is bounded away from 0 if and only if $W_2$ is bounded away from 0.

For iid observations, (\ref{hash.eg16})--(\ref{hash.eg20}) reduce to
\begin{eqnarray}
a_{10}
&=&
T(F),
\quad
a_{21} = [1^2] = W_2 [1^2]_1,
\nonumber
\\
a_{11}
&=&
[11]/2 = W_2 [11]_1 /2,
\nonumber
\\
a_{32}
&=&
[1^3] + 3[1,2,12] =  W_3 [1^3]_1  + 3 W_2^2[1,2,12]_{11},
\nonumber
\\
a_{22}
&=& [1,11] + [12^2]/2 + [1,122]
\nonumber
\\
&=&
W_3 [1,11]_1  +  W_2^2 [12^2]_{11}/2 +  W_2^2 [1,122]_{11},
\nonumber
\\
a_{43}
&=&
[1^4] - 3 W_4[1^2]_1^2
 +12[1,12,2^2] + 12[1,2,23,31]
 + 4[1,2,3,123]
\nonumber
\\
&=&
W_4 \{ \left[ 1^4\right]_1 -3\left[ 1^2 \right] ^2_1 \}
    + 12 W_2W_3 \left[ 1,12,2^2\right]_{11}
\nonumber
\\
&&
+ 4W_2^3 \{ 3 \left[ 1,2,23,31\right]_{111}
             +   \left[ 1,2,3,123\right]_{111} \},
\nonumber
\\
a_{12}
&=& \left[ 111\right]/6 + \left[ 1122\right]/8 =  W_3 \left[ 111\right]_1/6+  W_2^2 \left[ 1122\right]_{11}/8,
\nonumber
\end{eqnarray}
where $W_r$ is given by (\ref{Wr}).
This follows from the above results and (\ref{uniq}).
For $w_{in}\equiv 1$ these reduce to the expressions of Theorem 3.1 of Withers (1983).

\section{Two Simple Examples}
\addtocounter{section}{0}
\setcounter{equation}{0}

If $T(F)$ is a polynomial in $F$ of degree $r$, (for example, $\mu_r(F)$),
then derivatives of order greater than $r$ are zero.
Let us work through two simple examples: the mean and the variance.

\begin{example}
Suppose that $T(F)=\mu(F)=\mu$ say, and $s=1$.  Then $T_x=x-\mu$,
and higher derivatives are zero.
Then
\begin{eqnarray*}
a_{r,r-1}=n^{-1}\sum_{i=1}^n w_i^r\kappa_r(X_i)
\end{eqnarray*}
and other cumulant coefficients are 0.
If the observations are iid, then $a_{r,r-1}=W_r\kappa_r(X)$.

Now suppose that $\{w_i,F_i(x)\}$ can be parameterised as
$w_{in}=w(i/n)$, $F_{in}(x)=F(x,i/n)$
for some smooth functions $w(t),\ F(x,t)$. That is, $X_{in}$ has
distribution  $G_{\theta(i/n)}(x)$ say, and $ F(x,t)=G_{\theta(t)}(x)$.
Then we can write
\begin{eqnarray}
&&
E\ X_{in}^r = m_r(i/n), \mbox{ where }m_r(t)=\int x^rF(dx,t),
\nonumber
\\
&&
\kappa_r(X_{in}) = \kappa_r(i/n), \mbox{ where }
\kappa_1(t)=m_1(t),
\
\kappa_2(t)=m_2(t)-m(t)^2,\cdots
\nonumber
\\
&&
a_{r,r-1} = n^{-1}\sum_{i=1}^n k_r(i/n),
\mbox{ where }
k_r(t)=w(t)^r\kappa_r(t).
\label{kr}
\end{eqnarray}
That is, $m_r(t)$ and $\kappa_r(t)$ are the $r$th moment and cumulant of
$G_{\theta(t)}(x)$.
By the Euler-McLaurin expansion (Abramowitz and Stegun, 1964, Equation (23.1.30), page 806),
\begin{eqnarray*}
&&
n^{-1}\sum_{i=1}^{n}g(i/n) = \sum^{\infty}_{k=0} \alpha_{k} (g) n^{-k},
\mbox{ where}
\\
&&
\alpha_{0}(g) = \int_0^1 g(t) dt,
\nonumber
\\
&&
\alpha_{1}(g) = \left\{ g(1) - g(0) \right\} / 2,
\nonumber
\\
&&
\alpha_{k} (g) = \left\{ g^{(k-1)} (1) - g^{(k-1)} (0) \right\}
B_k / k!
\mbox{ for } k =2,3, \ldots
\nonumber
\end{eqnarray*}
and $B_k$ is the $k$th Bernoulli number, given by Abramowitz and Stegun (1964, page 809, last column):
$B_1=-1/2$, $B_2=1/6$,  $B_3=0$, $B_4=-1/30$, $\cdots$ and $B_k = 0$ for $k=3,5,7,\cdots$.
Applying this to $g=k_r$ of (\ref{kr}), we see that
the $r$th cumulant of  $\mu(\widehat{F})$ satisfies the basic expansion (\ref{hash.eg5}) with the new coefficients $a_{rj}' = \alpha_{j+1-r}(k_r)$.
So, the leading coefficients are
\begin{eqnarray*}
a_{r,r-1}' = \int_0^1 k_r(t)dt,
\
a_{rr}' =(k_r(1)-k_r(0))/2.
\end{eqnarray*}
In particular
\begin{eqnarray*}
a_{10}'=\int_0^1w(t)dt\int x F(dx,t),
\
a_{21}'=\int_0^1w(t)^2\kappa_2(t)dt.
\end{eqnarray*}
So, for the unweighted case $w(t)=1$,
\begin{eqnarray*}
a_{10}'=\int_0^1dt\int x F(dx,t),
\
a_{21}'=\int_0^1\kappa_2(t)dt.
\end{eqnarray*}

\noindent
{\bf Example 2.1.1} Suppose that $G_\theta(x)=1-e^{-x/\theta}$ on $(0,\infty)$,
the scaled exponential distribution. Then
\begin{eqnarray*}
&&
\kappa_r(t) = (r-1)!\theta(t)^r,
\
k_r(t)=(r-1)!\eta(t)^r, \mbox{where }\eta(t)=w(t)\theta(t),
\\
&&
a_{r,r-1}' =  (r-1)! \int_0^1\eta(t)^rdt,
\
a_{rr}' =(r-1)![\eta(1)^r-\eta(0)^r]/2.
\end{eqnarray*}
So, to this degree of approximation,
 weighting the observations amounts to weighting the scale parameter,
$\theta(t)$.
\end{example}

Since $\mu(F)$ is linear in $F$, the last example did not need the
machinery of functional differentiation. But as pointed out in Example 1.1,
this is not the case for the next example, the variance.

\begin{example}
Suppose that  $s=1$ and
$T(F)=\mu_2(F)=\mu_2$ say.
  Then $T_x=\mu_x^2-\mu_2$, where $\mu_x=x-\mu$,
$T_{xy}=-2\mu_x\mu_y$,
and higher derivatives are zero.
We give the cumulant coefficients in terms of
\begin{eqnarray}
M^k_{rs\cdots}=n^{-1}\sum_{i=1}^n w_i^k\mu_{ri}\mu_{si}\cdots,
\label{m}
\end{eqnarray}
where $\mu_{ri}=E \mu_{X_i}^r =E\  (X_i-\mu)^r$.
Since
\begin{eqnarray*}
\{1^r\}=n^{-1}\sum_{i=1}^n w_i^r\mu_{r}(\mu_{X_i}^2),
\end{eqnarray*}
we have $a_{21}=\{1^2\}=M^2_4-M^2_{22}$, $\{1^3\}=M^3_{6}-3M^3_{24}+2M^3_{222}$
and $\{1^4\}=M^4_{8}-4M^4_{26}+6M^4_{224}-3M^4_{2222}$.
Also
$[11]_i=-2E\ \mu_{X_i}^2=-2\mu_{2i}$, $[12]_{ij}=-2\mu_{1i}\mu_{1j}$, so that
\begin{eqnarray*}
a_{11}=-M_2^2 +M_{11}^2.
\end{eqnarray*}
Also
$[1]_i=\mu_{2i}-\mu_{2}$, $[1,12]_{ij}=-2(\mu_{3i}-\mu_2\mu_{1i})\mu_{1j}$,
$[1,2,12]_{ij}=-2(\mu_{3i}-\mu_2\mu_{1i})(\mu_{3j}-\mu_2\mu_{1j})$,
giving
$\{1,2,12\}=-2(M_3^2 -M_{12}^2)^2$,
so that
\begin{eqnarray*}
a_{32}=M^3_{6}-3M^3_{24}+2M^3_{222}-6(M_3^2-M_{12}^2)^2.
\end{eqnarray*}
Also $\{1,11\}=-2(M_4^3-2M_{13}^3-M_{22}^3+2M_{112}^3)$, $\{12^2\}=4(M_2^2-M_{11}^2)^2$, $\{1,122\}=0$, so that
\begin{eqnarray*}
a_{22}/2=(M_2^2-M_{11}^2)^2
-M^3_{4}+M^3_{22}+2M^3_{13}-2M_{112}^3.
\end{eqnarray*}
Similarly writing (\ref{hash.eg19}) as $a_{43}=b_1-3b_2+12b_3+12b_4+4b_5$, one obtains
\begin{eqnarray*}
b_1
&=&
M_8^4-4M_{26}^4+6M_{224}^4-3M_{2222}^4,
\\
b_2
&=&
M_{44}^4-2M_{224}^4+M_{2222}^4,
\\
b_3
&=&
2M_3^3(-M_5^3+M_{14}^3 +M_{23}^3 -4M_{122}^3 +\mu_2M^3_3 +5\mu_2^2M^3_1)
\\
&& +2M_{12}^2[2M_5^3 -M_{23}^3 +2M_{122}^3
  -3\mu_2(M^3_3+M_{12}^3) +3\mu_2^2M^3_1]
\\
&&
-2M_1^2\mu_2[M_5^3+M_{14}^3 -2M_{122}^3 -2\mu_2(M^3_3+M_{12}^3) +8\mu_2^2M^3_1],
\\
b_4
&=&
4(M_3^2-\mu_2M_1^2)(M_3^2-M_{12}^2)(M_2^2-M_{11}^2),
\\
b_5
&=&
0.
\end{eqnarray*}
Specialising to the iid case gives $\mu_{ri}=\mu_{r}$,
\begin{eqnarray*}
&&
[1^2]_1 = \mu_4-\mu_2^2,
\
[11]_1  =-2\mu_2,
\
[1^3]_1  =\mu_6-3\mu_2\mu_4+2\mu_2^3,
\\
&&
[1,2,12]_{11} = -2\mu_3^2,
\
[1,11]_1 = -2(\mu_4-\mu_2^2),
\\
&&
[12^2]_{11} = 4\mu_2^2,
\
[1^4]_1 = \mu_8-4\mu_2\mu_6+6\mu_2^2\mu_4-3\mu_2^4,
\\
&&
[ 1,12,2^2]_{11} =-2\mu_3(\mu_5-2\mu_2\mu_3),
\
[ 1,2,23,31]_{111} =  4\mu_2\mu_3^2,
\\
&&
[ 1,2,3,123]_{111} = [1,122]_{11} = [ 111]_1 = [ 1122]_{11}=0,
\end{eqnarray*}
so that
\begin{eqnarray*}
a_{21}
&=&
W_2(\mu_4-\mu_2^2),\ a_{11}= -W_2\mu_2,
\\
a_{32}
&=&
W_3(\mu_6-3\mu_2\mu_4+2\mu_2^3) -6W_2^2 \mu_3^2,
\\
a_{22}
&=&
-2W_3 (\mu_4-\mu_2^2)+  2W_2^2\mu_2^2,
\\
a_{43}
&=&  W_4[\mu_8-4\mu_2\mu_6+6\mu_2^2\mu_4-3\mu_2^4-3 (\mu_4-\mu_2^2)^2] -24 W_2W_3\mu_3(\mu_5-2\mu_2\mu_3)
\\
&&
+48W_2^3 \mu_2\mu_3^2.
\end{eqnarray*}
For the unweighted case, these cumulant coefficients $\{a_{ri}\}$,
reduce to those of Example 2
of Withers (1983).
\end{example}

\section{A Chain Rule for More Complex Examples}
\addtocounter{section}{0}
\setcounter{equation}{0}

When $T(F)$ is a function of moments (for example, the coefficient of
variation or correlation),  the following chain rule is very useful.

Suppose that $f:R^p\rightarrow R$ is a smooth function and that $S(F)$ is
a smooth functional in $R^p$.
For $1\leq a\leq p$, let
 $S_{a\cdot x_1\cdots x_r}$ denote the  $r$th derivatives of $S_a(F)$.
Then by an extension of the chain rule for differentiating a function of a
function, the first three derivatives of $T(F)=f(S(F))$ are
\begin{eqnarray*}
&&
T_x = f_a S_{ax},
\
T_{x_1x_2}=
f_{a}S_{a\cdot x_1x_2} + f_{ab} S_{a\cdot x_1}S_{b\cdot x_2},
\\
&&
T_{x_1x_2x_3} = f_{a}S_{a\cdot x_1x_2x_3}
+ f_{ab}\sum^3_{123} S_{a\cdot x_1}S_{b\cdot x_2x_3}
+ f_{abc}S_{a\cdot x_1}S_{b\cdot x_2}S_{c\cdot x_3},
\end{eqnarray*}
where $f_a = f_{a_1\cdots a_r}=\partial^r f(s)/\partial s_1\cdots\partial s_r|_{s=S(F)}$,
we use the tensor summation convention of {\it implicit summation of pairs}
of $a,b,\cdots$, and $\sum^3_{123}$ sums over all 3 permutations of 1,2,3 giving distinct terms.
Set
\begin{eqnarray*}
\nu_i^{ab\cdots} = E\ S_{aX_i}S_{bX_i}\cdots\mbox{ and }
M^k(ab\cdots,cd\cdots,\cdots)=n^{-1}\sum_{i=1}^n w_i^k
\nu_i^{ab\cdots}\nu_i^{cd\cdots}\cdots .
\end{eqnarray*}
Then
\begin{eqnarray*}
&&
[1]_i =
f_a  \nu^a_i,
\
[1^2]_i=
f_a f_b \nu^{ab}_i,
\\
&&
[1^3]_i = f_a f_b f_c \nu^{abc}_i,
\\
&&
[11]_i  = f_{ab}\nu^{ab}_i +
f_{a}E\ S_{aX_iX_i},
\\
&&
[12]_{ij} =  f_{ab} \nu^a_i \nu^b_j +
f_{a}E^{ind}\ S_{aX_iX_j},
\\
&&
[1,12]_{ij} =
f_{c} (f_{ab}\nu^{ac}_i \nu^b_j
+
f_{a} E^{ind}\  S_{cX_i}   S_{aX_iX_j} ),
\\
&&
[1,2,12]_{ij} = f_a f_b
(f_{cd}\nu^{ac}_i \nu^{bd}_j
+
f_{c} E^{ind}\  S_{aX_i}S_{bX_j}   S_{cX_iX_j} ).
\end{eqnarray*}
Set
\begin{eqnarray*}
&&
C(ab)  = \mu( S_{aX_i},S_{bX_i})
=M^2(ab)-M^2(a,b),
\\
&&
C(abc) =  n^{-1}\sum_{i=1}^n w_i^3
\mu( S_{aX_i},S_{bX_i},S_{cX_i})
=M^3(abc)-\sum^3_{abc} M^3(ab,c)+2M^3(a,b,c)
\end{eqnarray*}
since
\begin{eqnarray*}
\mu( S_{aX_i},S_{bX_i},S_{cX_i}) = \nu^{abc}_i-\sum^3_{abc}\nu^{ab}_i\nu^{c}_i
+2\nu^{a}_i\nu^{b}_i\nu^{c}_i.
\end{eqnarray*}
Then
\begin{eqnarray*}
&&
a_{21} = f_af_bC (ab),
\\
&&
2a_{11} =
f_{ab}C(ab) +
f_{a}\ n^{-1}\sum_{i=1}^n w_i^2
(E\ S_{aX_iX_i}-E^{ind}\ S_{aX_iX_j}|_{j=i}).
\end{eqnarray*}
Also $a_{32}$ is given by (\ref{hash.eg17}) in terms of
\begin{eqnarray*}
\{1^3\}
&=&
f_{a}f_bf_c C(abc),
\\
\{1,2,12\}
&=&
f_af_bf_c
\
n^{-2}\sum_{i,j=1}^n w_i^2w_j^2E^{ind}\ S_{aX_iX_j}\
(S_{bX_i}S_{cX_j}-2\nu_j^bS_{cX_i} +\nu_i^b\nu_j^c)
\\
&&
+f_af_bf_{cd} \ C(ad)C(bc).
\end{eqnarray*}
For the iid case these reduce to
\begin{eqnarray*}
&&
a_{21} = W_2f_af_b\nu^{ab}_1,
\\
&&
a_{11} = W_2(f_{ab}\nu^{ab}_1+f_aE\ S_{aXX})/2,
\\
&&
a_{32} = f_af_bf_c(W_3\nu^{abc}_1+3W_2^2E^{ind}\ S_{aX_1X_2} S_{bX_1} S_{cX_2})
+3f_af_bf_{cd} W_2^2\nu^{ad}_1\nu^{bc}_1.
\end{eqnarray*}

\begin{example}
Suppose that   $s=1,\ p=2,\ S_1(F)=\mu,\ S_2(F)=\mu_2$.
After some simplification one obtains
\begin{eqnarray*}
&&
a_{21} = f_af_bC (ab) = f_1^2C(11)+2f_1f_2C(12)+f_2^2C(22),
\\
&&
2a_{11} = f_{ab}C(ab) -2f_2 C(11)
= (f_{11}-2f_2) C(11) +2f_{12} C(12)+f_{22} C(22),
\\
&&
a_{32} =  \{ 1^3\} + 3 \{ 1, 2, 12\},
\end{eqnarray*}
where
\begin{eqnarray*}
&&
\{1^3\} = f_1^3C(111)+3f_1^2f_2C(112)+3f_1f_2^2C(122)+f_2^3C(222),
\\
&&
\{1,2,12\} =  D_1^2(f_{11}-f_2)+2D_1D_2f_{12} +D_2^2f_{12},
\
D_j = f_aC(ja).
\end{eqnarray*}
Also, in the notation of (\ref{m}), $C(ab),\ C(abc)$ reduce to
\begin{eqnarray*}
&&
C(ab) = M_{a+b}^2-M_{ab}^2;
\\
&&
C(abc) = M_{a+b+c}^2-\sum^3_{abc} M_{a+b,c}^2+2M_{a,b,c}^2:
\\
&&
C(111) = M_3^3,\ C(112)=M_4^3-M_{22}^3-2M_{13}^3+2M_{112}^3,
\\
&&
C(122) = M_5^3-M_{14}^3-2M_{23}^3+2M_{122}^3,
\
C(222)=M_6^3-3M_{24}^3+2M_{222}^3.
\end{eqnarray*}
For the iid case,
\begin{eqnarray*}
&&
C(11) = W_2\mu_2,\ C(12)=W_2\mu_3,\ C(22)=W_2(\mu_4-\mu_2^2),
\\
&&
C(111) = W_3\mu_3,
\
C(112)=W_3(\mu_4-\mu_2^2),
\\
&&
C(122) = W_3(\mu_5-2\mu_2\mu_3),
\
C(222)=W_3(\mu_6-3\mu_2\mu_4+2\mu_2^3).
\end{eqnarray*}
\end{example}

We now give two applications of this example, the Studentised
mean and the coefficient of variation.
Set $\sigma=\sigma(F)=\mu_2^{1/2}$ and $\lambda_r=\mu_r/\mu_2^{r/2}$.

\begin{example}
{\bf The Studentized mean.}
Take $s=1$, $T(G)=[\mu(G)-\mu(F)] /\sigma(G)$.
Then in terms of $C(ab),\ C(abc) $ of Example 2.4,
\begin{eqnarray*}
&&
T(F) = f_2=f_{11}=f_{22}=0,
\
f_1=\sigma^{-1},
\
f_{12}=-\sigma^{-3}/2,
\\
&&
a_{21} = C(11)/\mu_2,
\
a_{11}=-\sigma^{-3}C(12)/2,
\\
&&
a_{32} = C(111)\sigma^{-3}- 3C(11)C(12)\sigma^{-5}.
\end{eqnarray*}
For the iid case, $a_{21}=W_2$, $a_{11}=-W_2\lambda_3/2$, $a_{32}=(W_3-3W_2)^2 \lambda_3$.
For the unweighted case, this gives $a_{21}=1$, $a_{11}= -\lambda_3/2$, $a_{32}= -2 \lambda_3$,
in agreement with Withers (1989, Example 1.2, page 300).
\end{example}

\begin{example}
{\bf The inverse of the coefficient of variation.}
Suppose that  $s=1$ and
$T(F)=\mu/\sigma$.  Using the notation of the previous two examples, we obtain
\begin{eqnarray*}
&&
a_{21} = C(11)/\mu_2 - C(12)\mu/\mu_2^{2} +C(22)\mu^2/4\mu_2^{3},
\\
&&
2a_{11} = ( C(11)\mu-C(12))\sigma^{-3}
+3C(22)\mu\sigma^{-5}/4,
\\
&&
a_{32} =  \{ 1^3\} + 3 \{ 1, 2, 12\},
\end{eqnarray*}
where
\begin{eqnarray*}
&&
\{1^3\} = C(111)\sigma^{-3}-3C(112)\mu\sigma^{-5}/2
+3 C(122)\mu^2\sigma^{-7}/4 - C(222)\mu^3\sigma^{-9}/8,
\\
&&
\{1,2,12\} = (D_1^2\mu-D_1D_2)\sigma^{-3}
+3D_2^2\mu\sigma^{-5}/4,
\\
&&
D_1 = [C(11)-C(12)\mu/2\mu_2]\sigma^{-1},
\
D_2 =  [C(12)-C(22)\mu/2\mu_2]\sigma^{-1}.
\end{eqnarray*}
For the iid case,
\begin{eqnarray*}
&&
D_1  =  W_2\sigma\alpha_1, \mbox{ where }\alpha_1=1-\lambda_3 T(F)/2,
\\
&&
D_2  = W_2\sigma^2\alpha_2, \mbox{ where }
\alpha_2=\lambda_3-(\lambda_4-1) T(F)/2,
\end{eqnarray*}
and
\begin{eqnarray*}
&&
a_{21} = W_2[1-\lambda_3 T(F) +(\lambda_4-1) T(F)^2/4],
\\
&&
2a_{11} =  W_2[ T(F)(1+3\lambda_4)/4-\lambda_3],
\\
&&
\{1^3\}/W_3 =
\lambda_3 -3T(F)(\lambda_4-1)/2
 +3T(F)^2(\lambda_5-2\lambda_3)/4
-T(F)^3(\lambda_6-3\lambda_4+2)/8,
\\
&&
\{1,2,12\}/W_2^2 = \alpha_1^2T(F)-\alpha_1\alpha_2
+3\alpha_2^2T(F),
\\
&&
a_{32} = \{ 1^3\} + 3 \{ 1, 2, 12\}.
\end{eqnarray*}
\end{example}

\newpage

\begin{appendix}
\section*{Appendix A: Edgeworth-Cornish-Fisher Expansions}
\addtocounter{section}{1}
\setcounter{equation}{0}

Since the cumulants of $T(\widehat{F})$ satisfy the expansion (\ref{hash.eg5}),
it follows by Withers (1984) that its standardized version,
$Y_n = n^{1/2} a^{-1/2}_{21} \{ T(\widehat{F}) - T(F) \}$, has the Edgeworth-Cornish-Fisher expansions
\begin{eqnarray*}
&&
P_n(x) = P(Y_n \leq x)
\approx
\Phi(x) - \phi(x) \sum^\infty_{r=1}
           h_r(x) n^{-r/2},
\\
&&
P(|Y_n| \leq x)
\approx
2\Phi (x) - 1-2 \phi(x) \sum^\infty_{r=1}
h_{2r}(x) n^{-r},
\\
&&
\Phi^{-1} (P_n(x))
\approx
x-\sum^\infty_{r=1} f_r(x) n^{-r/2},
\\
&&
P^{-1}_n (\Phi(x))
\approx
x + \sum^\infty_{r=1} g_r(x) n^{-r/2},
\end{eqnarray*}
where $\Phi$ and $\phi$ are the distribution and
density of a unit normal random variable and
$\{ h_r, f_r, g_r \}$ are certain polynomials given by
(4.1) of Withers (1984) in terms of the
standardized cumulant coefficients
$A_{ri} = a^{-r/2}_{21} a_{ri}$.
In terms of the Hermite polynomial $H e_i (x) = \phi(x)^{-1} (-d/dx)^i \phi(x)$, the first few are given by
\begin{eqnarray*}
h_1
&=&
f_1 = g_1 = A_{11}
  + A_{32} He_2/6,
\\
h_2
&=&
(A^2_{11}
  + A_{22})H e_1/2
  + (4A_{11}A_{32} +A_{43}) H e_3/24
  + A^2_{32} H e_5/72,
\\
h_3
&=&
A_{12} + (A^3_{11} + 3A_{11}A_{22}+A_{33})H e_2/6
\\
& &
+ (10A^2_{11}A_{32} + 5A_{11}A_{43} + 10A_{22} A_{32}+A_{54})H e_1/120
\\
& &
+(2A_{11}A^2_{32} + A_{32}A_{43})H e_6/144 + A_{32} H e_8/1296,
\\
f_2(x)
&=&
(\ell_2/2-\ell_1\ell_3/3)x
  + \ell_4(x^3-3x)/24 - \ell^2_3 (4x^3 - 7x)/36,
\\
f_3(x)
&=&
-\ell_2\ell_2/2 + \ell^2_1\ell_3/6
    - \ell_2\ell_3 (5x^2-3)/12 - \ell_1 \ell_4(x^2-1)/8
\\
& &
+ \ell_5(x^4 - 6x^2 +3)/120
    + \ell_1\ell^2_3(12x^2 - 7)/36
    - \ell_3 \ell_4(11x^4 - 42x^2 + 15)/144
\\
& &
+ \ell^3_3 (69x^4 - 187 x^2 + 52)/648 + A_{12} + A_{33}      (x^2-1)/6,
\\
g_2(x)
&=&
\ell_2 x/2 + \ell_4 (x^3 -3x)/24
    - \ell^2_3(2x^3 - 5x)/36,
\\
g_3(x)
&=&
-\ell_2\ell_3 (x^2-1)/6
   + \ell_5 (x^4 - 6x^2 +3)/120
   - \ell_3 \ell_4 (x^4 - 5x^2 + 2)/24
\\
& &
+ \ell^3_3 (12x^4 - 53x^2 + 17) /324 + A_{12} + A_{33}(x^2-1)/6,
\end{eqnarray*}
where $\ell_1 = A_{11}$, $\ell_2 = A_{22}$, $\ell_3 = A_{32}$, $\ell_4 = A_{43}$ and $\ell_5 = A_{54}$.
The coefficients given in \S 2 give
$\{ h_r, f_r, g_r, 1\leq r \leq 2 \}$
and so give the distribution and quantiles of $Y_n$ to $O(n^{-3/2})$.
The same is true of its density, since we can write
\begin{eqnarray*}
h_r(x) = \sum^{}_{i=1} \{ h_{ri} He_i (x): r+i \mbox{ odd } \},
\end{eqnarray*}
so
\begin{eqnarray*}
P_n(x)
\approx
\Phi(x) - \sum^\infty_{r=1} n^{-r/2}
\sum^{3r-1}_{i=1} h_{ri} (-\partial / \partial x)^i \phi(x),
\end{eqnarray*}
giving density
\begin{eqnarray*}
p_n(x) \approx \phi(x) \{ 1 + \sum^\infty_{r=1} n^{- r/2} \overline{h}_r(x) \},
\end{eqnarray*}
where
\begin{eqnarray*}
\overline{h}_r(x) = \sum^{3r-1}_{i=1} \{ h_{ri} He_{i+1} (x) : r+i \mbox{ odd} \}.
\end{eqnarray*}
So, $\overline{h}_1 = A_{11}H e_1 + A_{32} H e_3 /6$,
$\overline{h}_2 = (A^2_{11} + A_{22}) H e_2/2 + (4A_{11}A_{32} + A_{43}) H e_4 /24 + A^2_{32} H e_6 /72$,
and so on.

Suppose $ T (F)$ is $q$-variate and the weights $p$-variate.
Then the distribution and
density of $Y_n = n^{1/2} \{ T ( \widehat{F}) - T (F) \}$
have the multivariate Edgeworth expansions for $x$ in $R^q$
\begin{eqnarray*}
P_n (x) = P (Y_n \leq x) \approx \sum_{r=0}^\infty n^{-r/2} \widetilde{P}_r ( -\partial / \partial x) \Phi_V (x)
\end{eqnarray*}
and
\begin{eqnarray*}
p_n (x) = (\partial /\partial x_1) \ldots (\partial/ \partial
x_q) P (Y_n \leq x) \approx \sum_{r=0}^\infty n^{- r/2} \widetilde{P}_r (-\partial/\partial x) \phi_V (x),
\end{eqnarray*}
where $ \Phi_V (x)$ and $ \phi_V (x)$ are the distribution
and density of $ {\mathcal N} (0, V)$,
$V = (a_1^{\alpha \beta})$,
\begin{eqnarray*}
&&
\widetilde{P}_0 (t) = 1,
\\
&&
\widetilde{P}_1 (t) = a_1^\alpha t_\alpha + a_1^{\alpha \beta \gamma} t_\alpha t_\beta t_\gamma /6,
\\
&&
\widetilde{P}_2 (t) = a_2^{\alpha \beta} t_\alpha t_\beta /2 + a_3^{\alpha \beta
\gamma \delta} t_\alpha t_\beta t_\gamma t_\delta /24 + \widetilde{P}_1 (t)^2,
\end{eqnarray*}
and so on, again using implicit summation over repeated pairs of indices.

\newpage

\section*{Appendix B: Proofs for Section 2}
\addtocounter{section}{1}

Here we derive the results of Section 2.
By the von Mises expansion (\ref{1.4}) and (\ref{uniq}),
\begin{eqnarray*}
T(\widehat{F}) - T(F) = \sum^\infty_{r=1} T_{(r)}/r!,
\end{eqnarray*}
where $T_{(r)} = T_r(\widehat{F}, F) = O_p(n^{-r/2})$ for $T(F)$ regular.
We may express the moments of $T(\widehat{F}) - T(F)$ in terms of
$\alpha_{r s \ldots} = E\ T_{(r)} T_{(s)} \ldots = O(n^{-\nu})$,
where $\nu = R/2$ for $R$ even, $\nu = (R+1)/2$ for $R$ odd and $R = r + s + \ldots$.
Since
\begin{eqnarray*}
T_{(r)} = n^{-r}\sum^n_{i_1, \ldots, i_r=1} w_{i_1} \ldots w_{i_r} \int \ldots \int T_{x_1\ldots x_r}  d I_{i_1 1} \ldots dI_{i_r r},
\end{eqnarray*}
where $I_{ir} = I(X_i \leq x_r) - F_i (x_r)$, to simplify $\alpha_{rs\ldots}$
we need expressions for $D_{i1\ldots r} = E\ I_{i1} \ldots I_{ir}$.
Set
\begin{eqnarray*}
&&
\delta_{ij} \ldots =I(i=j= \ldots ),
\\
&&
\delta_{ij\ldots, k\ell \ldots} =
I(i=j=\ldots \neq k = \ell = \ldots),
\\
&&
\delta_{ij,k\ell, mn} = I(i=j, k=\ell, m=n \mbox{ all three distinct}),
\end{eqnarray*}
and so on.
Then
\begin{eqnarray*}
&&
E\ I_{i1} I_{j2} = \delta_{ij} D_{i12},
\\
&&
E\ I_{i1} I_{j2} I_{k3} = \delta_{ijk} D_{i123},
\\
&&
E\ I_{i1} I_{j2} I_{k3} I_{\ell 4} =
\delta_{i\ldots \ell} D_{i1234}   + \sum^3 \delta_{ij, k\ell} D_{i 1 2} D_{k34}
\end{eqnarray*}
for
\begin{eqnarray*}
\sum^3 \delta_{ij,k\ell} D_{i 12} D_{k34}
&=&
\delta_{ij, k\ell} D_{i12} D_{k34}
  + \delta_{ik, j\ell} D_{i13} D_{j24}
+ \delta_{i\ell, jk} D_{i14} D_{j23},
\\
E\ I_{i1} \ldots I_{m5}
&=&
\delta _{i\ldots m} D_{i1\ldots 5}
   + \sum^{10} \delta_{ij, k\ell m} D_{i12} D_{k345},
\nonumber
\\
E\ I_{i1} \ldots I_{n6}
&=&
\delta_{i\ldots n} D_{i1\ldots 6}
  + \sum^{15} \delta_{ij, k\ldots n} D_{i12} D_{k3456}
\nonumber
\\
& & + \sum^{10} \delta_{ijk, \ell mn} D_{i123} D_{\ell 456}
   + \sum^{15} \delta_{ij, k\ell, mn} D_{i12} D_{k34} D_{m56}
\nonumber
\end{eqnarray*}
and $\sum^{10}$ and $\sum^{15}$ defined similarly.
Setting $F_{ir} = F_i(x_r)$ and $F_{i1 \wedge 2 \wedge \ldots} = F_i(\mbox{min}(x_1, x_2, \ldots ))$, we have
\begin{eqnarray*}
&&
D_{i1} =
0,
\\
&&
D_{i12} =
F_{i1\wedge 2} - F_{i1} F_{i2},
\\
&&
D_{i123} =
F_{i1\wedge 2 \wedge 3}
  - \sum^3_{123} F_{i1} F_{i2\wedge 3}
  + 2 F_{i1} F_{i2} F_{i3},
\\
&&
D_{i1\ldots 4} =
F_{i1\wedge \ldots \wedge 4}
- \sum^4_{1234} F_{i1} F_{i2\wedge 3 \wedge 4}
+ \sum^6_{1234} F_{i1} F_{i2} F_{i3\wedge 4}
- 3F_{i1} \ldots F_{i4},
\end{eqnarray*}
and so on, where $\sum^m_{1\ldots r} g_{1\ldots r} = \sum g_\pi$ summed over all
$m$ permutations of $1 \ldots r$ giving distinct terms.
So, writing $T_{1\ldots r} = T_{x_1\ldots x_r}$,
\begin{eqnarray*}
\alpha_1
&=&
E\ T_{(1)} = n^{-1} \sum_i w_i \int T_1 dD_{i1} = 0,
\\
\alpha_2
&=&
E\ T_{(2)} = n^{-2} \sum_{ij} w_iw_j \int\int T_{12} dE\ I_{i1} I_{j2}
\\
&=&
n^{-2} \sum_i w^2_i \int\int T_{12} dD_{i12}
\\
&=&
n^{-2} \sum w^2_i \{ \left[ 11\right]_i
   - \left[ 12 \right]_{ii} \}
  = n^{-1} \{ 11 \},
\\
\alpha_3
&=&
E\ T_{(3)} = n^{-3} \sum_{ijk} w_iw_jw_k \int\int\int T_{123}dE\ I_{i1} I_{j2} I_{k3}
\\
&=&
n^{-3} \sum_i w^3_i \int\int\int T_{123}dD_{i123}  = n^{-2} \{ 111 \},
\\
\alpha_4
&=&
E\ T_{(4)}  = n^{-4} \sum_{i \ldots \ell} w_i \ldots w_\ell
  \int \ldots \int dE\  I_{i1}\ldots I_{\ell 4}
\\
&=&
n^{-3} \alpha_{4\cdot1} + 3n^{-2} \alpha_{4\cdot 2}
\end{eqnarray*}
for
\begin{eqnarray*}
\alpha_{4\cdot 1} = n^{-1} \sum_i w_i^4 \int \ldots \int T_{1 \ldots 4} dD_{i1 \ldots 4} = O(1)
\end{eqnarray*}
and
\begin{eqnarray*}
\alpha_{4 \cdot 2} = n^{-2} \sum^\prime_{ik} w_i^2 w^2_k \int\ldots \int T_{1\ldots 4} dD_{i1 2} dD_{k34},
\end{eqnarray*}
where $\sum^\prime_{ij\ldots}$ sums over distinct $i,j \ldots$ in $1, \ldots, n$.
So, $\alpha_{4.2} = \alpha^\prime_{4.2} + O(n^{-1})$, where
$\alpha^\prime_{4.2}$ replaces $\sum^\prime_{ik}$ by $\sum_{ik}$.
So, $\alpha^\prime_{4.2} = \{ 1122\}$.
The expressions for $a_{10}, a_{11}, a_{12}$ follow since $a^\prime_{1i} = a_{1i}$.
Also
\begin{eqnarray*}
E\  \{T(\widehat{F})
- T(F)\}^2
= \alpha_{11}
+ \alpha_{12}
+ \alpha_{13}/3
+ \alpha_{22}/4
+ O(n^{-3}),
\end{eqnarray*}
where
\begin{eqnarray*}
\alpha_{11}
&=&
n^{-2} \sum_{ij} w_iw_j \int\int T_1T_2 dE\ I_{i1}I_{j2}
  = n^{-2} \sum_i w^2_i \int\int T_1 T_2 dD_{i12} \\
&=& n^{-1} \{ 1^2 \},
\\
\alpha_{12}
&=&
n^{-3} \sum_{ijk} w_i w_jw_k \int\int\int T_1 T_{23}
   dE\ I_{i1}I_{j2} I_{k3}
\\
&=&
n^{-3} \sum_i w^3_i \int\int\int T_1 T_{23} dD_{i123}
  = n^{-2} \{ 1, 11 \},
\\
\alpha_{13}
&=&
n^{-4} \sum_{i\ldots \ell} w_i \ldots w_\ell \int\ldots\int
T_1T_{234} dE\ I_{i1}\ldots I_{\ell 4} = \alpha_{13.1} + \alpha_{13.2},
\\
\alpha_{22}
&=&
\alpha_{22.1} + \alpha_{22.2}
\end{eqnarray*}
for
\begin{eqnarray*}
&&
\alpha_{13.1} = n^{-4} \sum_i w^4_i \int\ldots\int T_1 T_{234} dD_{i1234} = O(n^{-3}),
\\
&&
\alpha_{13.2} = n^{-4} \sum^\prime_{ik} w^2_i w^2_k \int\ldots \int T_1T_{234} d(D_{i12} D_{k34}) = \alpha^\prime_{13.2} + O(n^{-3}),
\\
&&
\alpha_{22.1} = n^{-4} \sum_i w^4_i \int\ldots\int T_{12}T_{34} dD_{i1\ldots 4} = O(n^{-3}),
\\
&&
\alpha_{22.2} = \alpha_{22.21} + 2\alpha_{22.22} + O(n^{-3}),
\\
&&
\alpha_{22.21} = n^{-4} \sum_{ij}w^2_i w^2_j\int\int T_{12}T_{34} d(D_{i12} D_{j34}) = n^{-2} \{ 11 \}^2,
\\
&&
\alpha_{22.22} = n^{-4} \sum_{ij}w^2_i w^2_j \int \int T_{12}T_{34} d(D_{i13} D_{j24}) = n^{-2} \{ 12 ^2\},
\end{eqnarray*}
where $\alpha^\prime_{13.2}$ replaces $\sum^\prime_{ik}$ by $\sum_{ik}$, so $\alpha^\prime_{13.2} = n^{-2} \{ 1, 122 \}$.
So, $a^\prime_{21} = \{ 1^2\}$ and $a^\prime_{22} = \{ 1, 11 \} + \{ 1, 122 \} + \{ 11 \}^2/4 + \{ 12^2 \} /2$.
Now use $a_{21} = a^\prime_{21}$ and
$a_{22} = a^\prime_{22} - a^2_{11}$.
Also
\begin{eqnarray*}
E\  \{T(\widehat{F}) - T(F)\}^{3} = \alpha_{111} + 3\alpha_{112}/2 + O(n^{-3}),
\end{eqnarray*}
where
\begin{eqnarray*}
\alpha_{111}
&=&
n^{-3} \sum_{ijk} w_iw_jw_k\int\int\int T_1T_2T_3
dE\ I_{i1} I_{j2} I_{k3}
\\
&=&
n^{-3} \sum_i w^3_i \int\int\int T_1T_2T_3 dD_{i123}
  = n^{-2} \{ 1^3\}
\end{eqnarray*}
and
\begin{eqnarray*}
\alpha_{112}
&=&
n^{-4} \sum_{i\ldots \ell} w_{i} \ldots w_{\ell}
\int \ldots \int T_1T_2T_{34} dE\ I_{i1} \ldots I_{\ell 4}
\\
&=&
\alpha_{112.1} + \alpha_{112.2} + 2\alpha_{112.3}
\end{eqnarray*}
for
\begin{eqnarray*}
\alpha_{112.1}
&=&
n^{-4} \sum_i w^4_i \int\ldots \int T_1T_2T_3 dD_{i1\ldots 4} = O(n^{-3}),
\\
\alpha_{112.2}
&=&
n^{-4} \sum^\prime_{ij} w^2_iw^2_j \int\ldots\int T_1T_2T_{34} d(D_{i12} D_{j34})
\\
&=& n^{-2} \{ 1^2 \} \{ 11 \}  + O(n^{-3}),
\\
\alpha_{112.3}
&=&
n^{-4} \sum^\prime_{ij} w^2_i w^2_j \int\ldots\int T_1T_2T_{34} d(D_{i13} D_{j24})
\\
&=&
n^{-2} \{ 1, 2, 12 \} + O(n^{-3}).
\end{eqnarray*}
So, $a^\prime_{32} = \{ 1^3 \}  + 3 \{ 1^2 \} \{ 11 \} /2  + 3 \{ 1, 2, 12 \}$.
Now use $a_{32} = a^\prime_{32} - 3a_{21} a_{11}$ to obtain $a_{32}$ above.
Also
\begin{eqnarray*}
&&
E\  \{ T(\widehat{F}) - T(F) \}^4 = \alpha_{1111}  + 2\alpha_{1112}  + 2\alpha_{1113}/3  + 3\alpha_{1122}/2  + O(n^{-4}),
\\
&&
\alpha_{1111} = \alpha_{1111.1} + 3\alpha_{1111.2},
\\
&&
\alpha_{1112} = \alpha_{1112.1}
  + 3\alpha_{1112.2}
  + 6\alpha_{1112.3}
  + \alpha_{1112.4},
\\
&&
\alpha_{1113} = \alpha_{1113.1} + O(n^{-4}),
\\
&&
\alpha_{1122} = \alpha_{1122.1} + O(n^{-4})
\end{eqnarray*}
for
\begin{eqnarray*}
\alpha_{1111.1}
&=&
n^{-4} \sum_i w^4_i \int\ldots\int T_1T_2T_3T_4 dD_{1234} = n^{-3} \{ 1^4 \},
\\
\alpha_{1111.2}
&=&
n^{-4}\sum^\prime_{ij} w^2_i w^2_j
\int\ldots\int T_1\ldots T_4  d(D_{i12} D_{j34})
\\
&=&
n^{-2} \{ 1^2 \}^2 - n^{-3} \{ 1^2, 1^2 \},
\\
\alpha_{1112.1}
&=&
n^{-5} \sum_iw^5_i \int\ldots \int T_1T_2T_3T_{45} dD_{i1\ldots 5} = O(n^{-4}),
\\
\alpha_{1112.2}
&=&
n^{-5} \sum^\prime_{ij}w^2_i w^3_j \int\ldots \int  T_1T_2T_3T_{45} d(D_{i12} D_{j345})
\\
&=&
n^{-3} \{ 1^2 \} \{ 1, 11 \}  + O(n^{-4}),
\\
\alpha_{1112.3}
&=&
n^{-5} \sum^\prime_{ij} w^2_i w^3_j \int\ldots\int T_1T_2T_3T_{45} d(D_{i14} D_{j235})
\\
&=&
n^{-3} \{ 1, 12, 2^2 \} + O(n^{-4}),
\\
\alpha_{1112.4}
&=&
n^{-5} \sum^\prime_{ij} w^2_i w^3_j \int\ldots \int
T_1T_2T_3T_{45} d(D_{i45} D_{j123})
\\
&=&
n^{-3} \{ 1^3 \} \{ 11 \} + O(n^{-4}),
\\
\alpha_{1113.1}
&=&
n^{-6} \sum_{ijk} w^2_i w^2_j w^2_k
\int\ldots\int T_1T_2T_3 T_{456} d\sum^{15}
D_{i12} D_{j34} D_{k56}
\\
&=&
9 \alpha_{1113.2} + 6\alpha_{1113.3},
\\
\alpha_{1113.2}
&=&
n^{-6} \sum_{ijk} w^2_i w^2_j w^2_k \int\ldots\int
T_1T_2T_3T_{456} d(D_{i12} D_{j34} D_{k56})
\\
&=&
n^{-3} \{ 1^2\} \{ 1, 122 \},
\\
\alpha_{1113.3}
&=&
n^{-6} \sum_{ijk} w^2_i w^2_jw^2_k \int\ldots\int
T_1T_2T_3T_{456} d(D_{i14} D_{j25}D_{k36})
\\
&=&
n^{-3} \{ 1, 2, 3, 123 \},
\\
\alpha_{1122.1}
&=&
n^{-6} \sum_{ijk} w^2_i w^2_j w^2_k
\int\ldots \int T_1T_2T_{34}T_{56} d \sum^{15}
D_{i12} D_{j34} D_{k56}
\\
&=&
n^{-3} (\gamma_1 + 4 \gamma_2 + 8\gamma_3 + 2\gamma_4),
\end{eqnarray*}
where
\begin{eqnarray*}
\gamma_1
&=&
n^{-3} \sum_{ijk} w^2_i w^2_j w^2_k \int\ldots \int T_1 T_2 T_{34}
T_{56} d(D_{i12} D_{j34} D_{k56})
\\
&=&
\{ 1^2 \} \{ 11 \}^2,
\\
\gamma_2
&=&
n^{-3} \sum_{ijk} w^2_i w^2_j w^2_k \int\ldots\int
T_1 T_2 T_{34} T_{56} d(D_{i13} D_{j24} D_{k56})
\\
&=&
\{ 11 \} \{ 1, 2, 12 \},
\\
\gamma_3
&=&
n^{-3}\sum_{ijk} w^2_i w_j^2 w_k^2 \int\ldots \int T_1 T_2 T_{34}
T_{56} d(D_{i13} D_{j45} D_{k26})
\\
&=&
\{ 1, 13, 32, 2 \},
\\
\gamma_4
&=&
n^{-3} \sum_{ijk}w_i^2 w_j^2 w_k^2  \int \ldots \int T_1 T_2
T_{34} T_{56} d(D_{i12} D_{j35} D_{k46})
\\
&=&
\{ 1^2\} \{ 12^2 \}.
\end{eqnarray*}
So, $a^\prime_{42} = 3 \{ 1^2 \}^2$ and $a^\prime_{43} = a_{43} + 4a_{11} a_{32} + 6a_{21} a^\prime_{22}$ for $a_{43}$ as above.
Now use $a_{43} = a^\prime_{43} - 4a_{11}a_{32} - 6a_{21}a^\prime_{22}$.

\begin{note}
An alternative method better for obtaining these expressions for the
cumulant coefficients $a_{ri}$,
 is to use the parametric approach of Withers (1988).
This is possible since
\begin{eqnarray*}
\kappa (\widehat{F}(x_1), \ldots, \widehat{F}(x_r)) = n^{1-r}  k  (x_1\ldots x_r),
\end{eqnarray*}
where
\begin{eqnarray*}
k(x_1 \ldots x_r) = n^{-1} \sum^r_{i=1} w^r_i k (x_1\ldots x_r F_i)
\end{eqnarray*}
and $k (x_1 \ldots x_r F_i) = \kappa (I(X_i \leq x_1), \ldots, I (X_i  \leq x_r))$.
\end{note}

\end{appendix}


\begin{thebibliography}{999}

\bibitem{}
Abramowitz, M. and Stegun, I. A. (1964).
{\it Handbook of Mathematical Functions}.
National Bureau of Standards, Washington D.C.

\bibitem{}
Chernoff, H. and Zacks, S. (1964).
Estimating the current mean of a normal distribution which is subjected to changes in time.
{\it Annals of Mathematical Statistics}, {\bf 35}, 999--1018.

\bibitem{}
Hajek, J. and Sidak, Z. (1967)
{\it Theory of Rank Tests},
Academic Press, Inc., New York.

\bibitem{}
Jaeckel, L. A. (1972).
The infinitesmal jackknife.
Bell Laboratories Technical Report MM 72-1215-11, June 30, 1972, New Jersey.

\bibitem{}
Kander, Z. and Zacks, S. (1966).
Test procedures for possible changes in parameters of statistical distributions occurring at unknown time points.
{\it Annals of Mathematical Statistics}, {\bf 37}, 1196--1210.

\bibitem{}
Koul, H. L. (1992).
Weighted empiricals and linear models.
{\it Institute of Mathematical Statistics Monograph Series}, {\bf 21}, Michigan State University.

\bibitem{}
Koul, H. L. (2002).
Weighted empirical processes in dynamic nonlinear models. Second edition.
{\it Lecture Notes in  Statistics}, {\bf 166}, Springer, New York.

\bibitem{}
Lahiri, S. N. (1992a).
On bootstrapping M-estimators.
{\it Sankhy\=a}, A, {\bf 54}, 157--170.

\bibitem{}
Lahiri, S. N. (1992b).
Edgeworth expansions for M-estimators of a linear regression parameter.
{\it Journal of Multivariate Analysis}, {\bf 43}, 125--132.


\bibitem{}
Lahiri, S. N. (1992c).
Bootstrapping M-estimators of a multiple linear regression parameter.
{\it Annals of Statistics}, {\bf 20}, 1548--1570.


\bibitem{}
Lahiri, S. N. (2003).
{\it Resampling Methods for Dependent Data}.
Springer, New York.


\bibitem{}
Withers, C. S. (1983).
Expansions for the distribution and quantiles of a regular functional of the empirical distribution
with applications to nonparametric condidence intervals.
{\it Annals of Statistics}, {\bf 11}, 577--587.



\bibitem{}
Withers, C. S. (1988).
Nonparametric confidence intervals for functions of several distributions.
{\it Annals of the Institute of Statistical Mathematics}, {\bf 40}, 727--746.

\bibitem{}
Withers, C. S. (1989).
The distribution and cumulants of a Studentised statistic.
{\it Communications in Statistics---Theory and Methods}, {\bf 18}, 295--318.


\bibitem{}
Withers, C. S. and Nadarajah, S. (2008).
Edgeworth expansions for functions of weighted empirical distributions
with applications to nonparametric confidence intervals.
{\it Journal of Nonparametric Statistics}, {\bf 20}, 751--768.



\bibitem{}
von Mises, R. (1947).
On the asymptotic distribution of differentiable statistical functionals.
{\it Annals of Mathematical Statistics}, {\bf 18}, 309--348.


\end{thebibliography}
\end{document}